\documentclass{article}
\usepackage{spconf,amsmath,graphicx}
\usepackage{algorithm}
\usepackage{algorithmic}
\usepackage{booktabs}
\usepackage{microtype}
\usepackage{graphicx}
\usepackage{subfigure}
\usepackage{hyperref}
\usepackage{multirow}
\usepackage{color}
\usepackage{amsthm,amsmath,amssymb}
\usepackage{mathrsfs}

\title{Multi-QuartzNet: Multi-Resolution Convolution for Speech Recognition with Multi-Layer Feature Fusion}
\name{Jian Luo, Jianzong Wang*\thanks{*Corresponding author: Jianzong Wang, jzwang@188.com}, Ning Cheng, Guilin Jiang, Jing Xiao}
\address{Ping An Technology (Shenzhen) Co., Ltd.}
\begin{document}
%
\maketitle
\begin{abstract}
In this paper, we propose an end-to-end speech recognition network based on Nvidia's previous QuartzNet \cite{KrimanQuartzNet} model. We try to promote the model performance, and design three components: (1) Multi-Resolution Convolution Module, replaces the original 1D time-channel separable convolution with multi-stream convolutions. Each stream has a unique dilated stride on convolutional operations. (2) Channel-Wise Attention Module, calculates the attention weight of each convolutional stream by spatial channel-wise pooling. (3) Multi-Layer Feature Fusion Module, reweights each convolutional block by global multi-layer feature maps. Our experiments demonstrate that Multi-QuartzNet model achieves CER $6.77\%$ on AISHELL-1 data set, which outperforms original QuartzNet and is close to state-of-art result.
\end{abstract}
\section{Introduction}
\label{sec:intro}

In the last few years, end-to-end neural networks have achieved remarkable results on automatic speech recognition tasks. Among these models, convolutional neural network architectures have attracted much attention. They are often used in the models combined with recurrent layers, such as CLDNN \cite{Sainath2015Convolutional}, DeepSpeech \cite{Hannun2014Deep, AmodeiDeepSpeech2}. In these works, the CNN layers are designed to reduce the time and spectral variation of the input features,  and their outputs are passed to RNN layers for temporal modeling. However, these models are often encountered with speed problems, because RNN layers cannot be trained or inferenced parallelly. Given the evidence that convolutional networks are also suitable on long-range dependency tasks, a lot of fully convolutional approachs were proposed. Wav2Letter \cite{CollobertWav2Letter} proposed an ASR system, which only used a standard 1D convolutional neural network trained by CTC loss. And then, fully convolutional networks \cite{ZeghidourFully} were presented. In their works, not only acoustic models but also language models as well as learnable front end were all based on convolutional pipelines.

Inspired by Wav2Letter, Nvidia's team proposed a computationally efficient end-to-end convolutional network named Jasper \cite{LiJasper}, which used a stack of 1D-convolution layers, with ReLU and batch normalization. They also found that simple ReLU and batch normalization outperform other activation and normalization. In Jasper, they introduced dense residual connections for training converge and better performance. They then updated their model to QuartzNet \cite{KrimanQuartzNet}, replacing traditional 1D-convolution layers by 1D time-channel separable convolutional layers. Time-depth separable convolutions \cite{HannunSequence, KumawatDepthwise, RahimianXceptionTime} are designed to reduce the number of parameters in traditional convolutions while keeping the receptive field large. The original QuartzNet model has $\mathbb{K}\times \mathbb{C}+\mathbb{C}^2$ parameters, where $\mathbb{K}$ is the kernel size and $\mathbb{C}$ is the channel dimension. By comparison, our proposed multi-resolution separable convolution has $\mathbb{K}\times \mathbb{C} \times \mathbb{S}+\mathbb{C}^2$ parameters, where $\mathbb{S}$ are the stream numbers of multi-resolution. The parameters of our model increase slightly, because stream numbers $S$ are usually quite small (e.g., $\mathbb{S}=2, 4$), but still dramatically better than traditional convolutions' $\mathbb{K}\times \mathbb{C}^2$ parameters. As shown in experimental section, multi-resolution convolutions get better results in our settings.

Some previous works have tried multi-stream or multi-resolution methods in their models. \cite{KyuMulti} proposed a multi-stride self-attention mechanism with various strides over neighboring speech frames. \cite{li2019multistream, li2018multiencoder} used a multi-stream idea combining CNN-based and RNN-based encoders. \cite{rownicka2019multiscale} proposed a multi-scale octave convolution layer to learn robust speech representations efficiently. Inspired from the work \cite{han2019stateoftheart}, we propose multi-resolution separable convolution with channel-wise attention. In their works, they used multi-stream for more effective expressions of speech features to the subsquent self-attention module. However, their model just concatenated all of the stream outputs to one vector, and did not consider the degree of importance of each convolutional stream. In our works, we design a channel-wise attention module, to calculate the attention weight of each convolutional stream automatically. The attention weight module is composed of squeeze and excitation \cite{hu2017squeezeandexcitation, woo2018cbam} function by pooling the feature maps across each channel dimension.

Another point of our works is that, we try to make connections cross the bottom and up layers. The bottom layers often contain local and detail information, and up layers by contrast usually contain global and general information. We think it is useful to combine them together for better speech expressions. Jasper \cite{LiJasper} designed Dense Residual (DR) \cite{huang2016densely}, making connections between the blocks. Feature Pyramid Networks (FPN) \cite{lin2016feature, guo2019augfpn} is another popular feature fusion mechanism used in the area of image object detection. FPN adoptes a backbone model, and builds feature pyramid. The feature pyramid sequentially combines two adjacent layers in feature hierarchy with top-down and lateral connections. Recently, Cross-layer Feature Pyramid Network (CFPN) \cite{li2020crosslayer} was proposed to improve the progressive fusion between layer-level information of the pyramid network. We adopt the Cross-layer Feature Aggregation (CFA) module of CFPN, aggregated multi-scale features from different convolutional blocks to a global context. The context is used to calculate the attention weight of each block. We design a multi-layer feature fusion module, which combines the re-weighted blocks with local residual connection. 

We demonstrate that the proposed Multi-QuartzNet model performs best, when using all of the three components: (1) Multi-Resolution Convolution Module, (2) Channel-Wise Attention Module, and (3) Multi-Layer Feature Fusion Module.

\section{Model Architecture}
\label{sec:majhead}

\begin{figure}[t]
	\begin{center}
		\centerline{\includegraphics[width=\columnwidth]{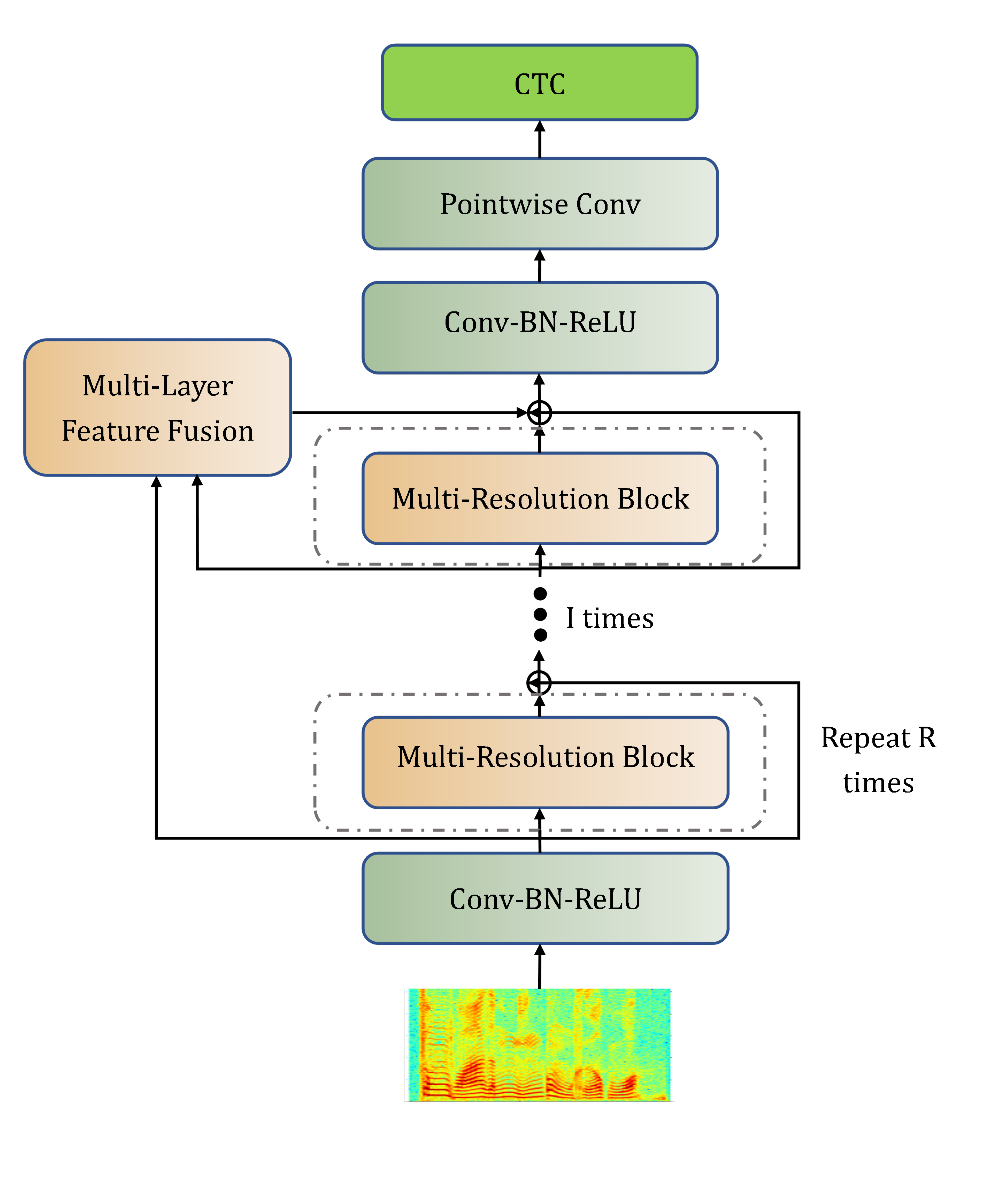}}
		\caption{Multi-QuartzNet Model Architecture}
		\label{fig1}
	\end{center}
	\vskip -0.2in
\end{figure}

Our proposed Multi-QuartzNet is designed based on QuartzNet architecture, which is a fully convolutional network trained on CTC loss. The model architecture of Multi-QuartzNet is shown in Figure~\ref{fig1}, which has the following structure: 1) The raw acoustic features are first fed into a 1D convolution layer $C_1$(Conv-BN-ReLU). 2) After that, it is followed by a sequence of multi-resolution blocks, and each block $B_i(i=1,...,\mathbb{I})$ is repeated $\mathbb{R}$ times. The block consists of $\mathbb{M}$ times multi-resolution convolution modules. Each convolution module is composed of a multi-resolution convolutional layer (DepthwiseConv-PointwiseConv-BN-ReLU) with stride set $[1,...,\mathbb{S}]$ and a channel-wise attention module to calculate the attention weight of each stream. 3) Morever, the model introduces a multi-layer feature fusion module, to reweight each convolutional block output. 4) Finally, three additional convolutional layers ($C_2$, $C_3$, $C_4$) are designed to map the features to vocabulary size with CTC loss training.

\begin{figure}[t]
	\begin{center}
		\centerline{\includegraphics[width=\columnwidth]{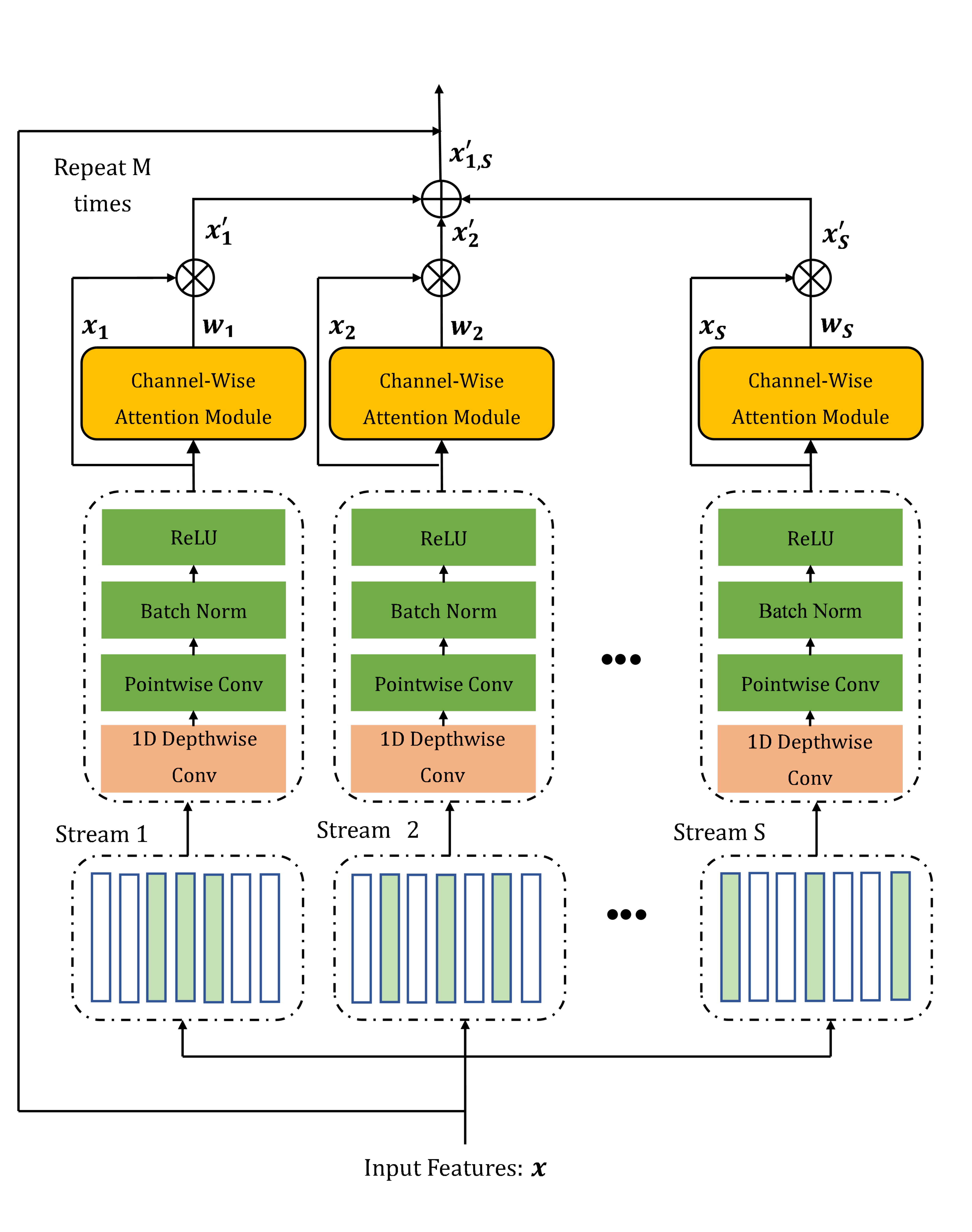}}
		\caption{Multi-Resolution Convolution Module}
		\label{fig2}
	\end{center}
\end{figure}

\subsection{Multi-Resolution Convolution}
\label{ssec:subhead}

Compared with the original QuartzNet architecture, we replace the 1D time-channel separable convolution with our proposed multi-resolution convolutions. Each convolution has a unique dilated stride but still time-channel separable. More specifically, we define an input feature map $x\in\mathbb{T}\times \mathbb{C}$ of each convolution, where $\mathbb{T}$ is the total frames number and $\mathbb{C}$ is the channel demension. For each stream, the feature $x$ is firstly opearated on a 1D-Depthwise convolutional layer with same kernel size $\mathbb{K}$ but with different stride $\mathbb{S}$. Because the depthwise convolutional layer fixes the channel demension and operates at spatial demension, each stream produces an output $x_\mathbb{S}$ with the same shape $\mathbb{T}\times \mathbb{C}$. Then the $x_\mathbb{S}$ is reviewed by the shape of $1\times \mathbb{T}\mathbb{C}$, and is applied by a $1 \times 1$ pointwise convolutions across all the channels. And then, batch norm layer and ReLU activation are attached after convolutional layer of each stream. After multi-resolution convolutions, the model sums all the stream outputs by channel dimension,
\begin{equation}
\label{sum_stream}
x_{1,\mathbb{S}} = x_1\oplus x_2...\oplus x_\mathbb{S}
\end{equation}
where $\oplus$ denotes the add operation. We maintain the same channel dimension to all the streams, so $x_{1,\mathbb{S}}$ has the same shape with input $x$, which is $x_{1,\mathbb{S}}\in \mathbb{T}\times \mathbb{C}$. The whole multi-resolution convolution module is shown as Figure~\ref{fig2}.

The multi-resolution convolution module enables the network to capture multi-scale information. With small dilated stride (e.g., $\mathbb{S}=1$), the stream will more focus on the local information like spectral of phoneme. With large dilated stride (e.g., $\mathbb{S}=4$), the stream will pay more attention to the global information like background noise. By contrast, traditional convolution just looks around the local information which is restricted by kernel size $\mathbb{K}$.

\subsection{Channel-Wise Attention Module}
\label{ssec:subhead}

As Figure~\ref{fig2} depicts, we design a channel-wise attention module in multi-resolution convolution. The attention module uses squeeze-and-excitation function to calculate the attention weight of each channel. The convolutional outputs $x_\mathbb{S}$ are then reweighted by multiplying the attention weight across each channel dimension. Inspired by SENet \cite{hu2017squeezeandexcitation}, the channel-wise attention module firstly squeezes spatial information into a channel-wise vector by average and maximum pooling. Formally, the average pooling vector $x_\mathbb{S}^{avg}\in \mathbb{C}$ and max pooling vector $x_\mathbb{S}^{max}\in \mathbb{C}$, are calculated by:
\begin{equation}
x_\mathbb{S}^{avg} = F^{avg}(x_\mathbb{S}) = \frac{1}{\mathbb{T}}\sum_{t=1}^\mathbb{T}x_\mathbb{S}(t)
\end{equation}
\begin{equation}
x_\mathbb{S}^{max} = F^{max}(x_\mathbb{S}) = \max_{t=1}^\mathbb{T}x_\mathbb{S}(t)
\end{equation}
Here squeezed vector $x_\mathbb{S}^{avg}$ and $x_\mathbb{S}^{max}$ are calculated separately by each channel. Next, the excitation function takes the squeezed vector into a gating mechanism network, which is composed of two fully connected layers with a sigmoid activation:
\begin{equation}
\begin{split}
w_\mathbb{S} = F^{ex}(x_\mathbb{S}^{avg}, x_\mathbb{S}^{max}) = \sigma(g(x_\mathbb{S}^{avg}, x_\mathbb{S}^{max}))\\
= \sigma(w_2\delta(w_1x_\mathbb{S}^{avg}) \oplus w_2\delta(w_1x_\mathbb{S}^{max}))
\end{split}
\end{equation}
where $\delta$ refers to the ReLU activation, and $\sigma$ denotes the sigmoid function. $w_1\in\mathbb{C}/\mathbb{D}\times\mathbb{C}$ and $w_2\in\mathbb{C}\times\mathbb{C}/\mathbb{D}$, are the gating network weights. $\mathbb{D}$ is the dimension reduction ratio. So output $x^\prime_\mathbb{S}$ of each convolution are reweighted by channel-wise product of $x_\mathbb{S}$ and $w_\mathbb{S}$. After that, all of the reweighted streams are summed together like Eq.~\ref{sum_stream}:
\begin{equation}
x^\prime_\mathbb{S} = F^{scale}(x_\mathbb{S}, w_\mathbb{S}) = x_\mathbb{S}*w_\mathbb{S}
\end{equation}
\begin{equation}
x^\prime_{1,\mathbb{S}} = x^\prime_1\oplus x^\prime_2...\oplus x^\prime_\mathbb{S}
\end{equation}
The channel-wise attention module gives attention values to each stream, making the network excite and suppress the stream referring to whole spatial information.

\subsection{Multi-Layer Feature Fusion}
\label{ssec:subhead}

\begin{figure}[t]
	\begin{center}
		\centerline{\includegraphics[width=\columnwidth]{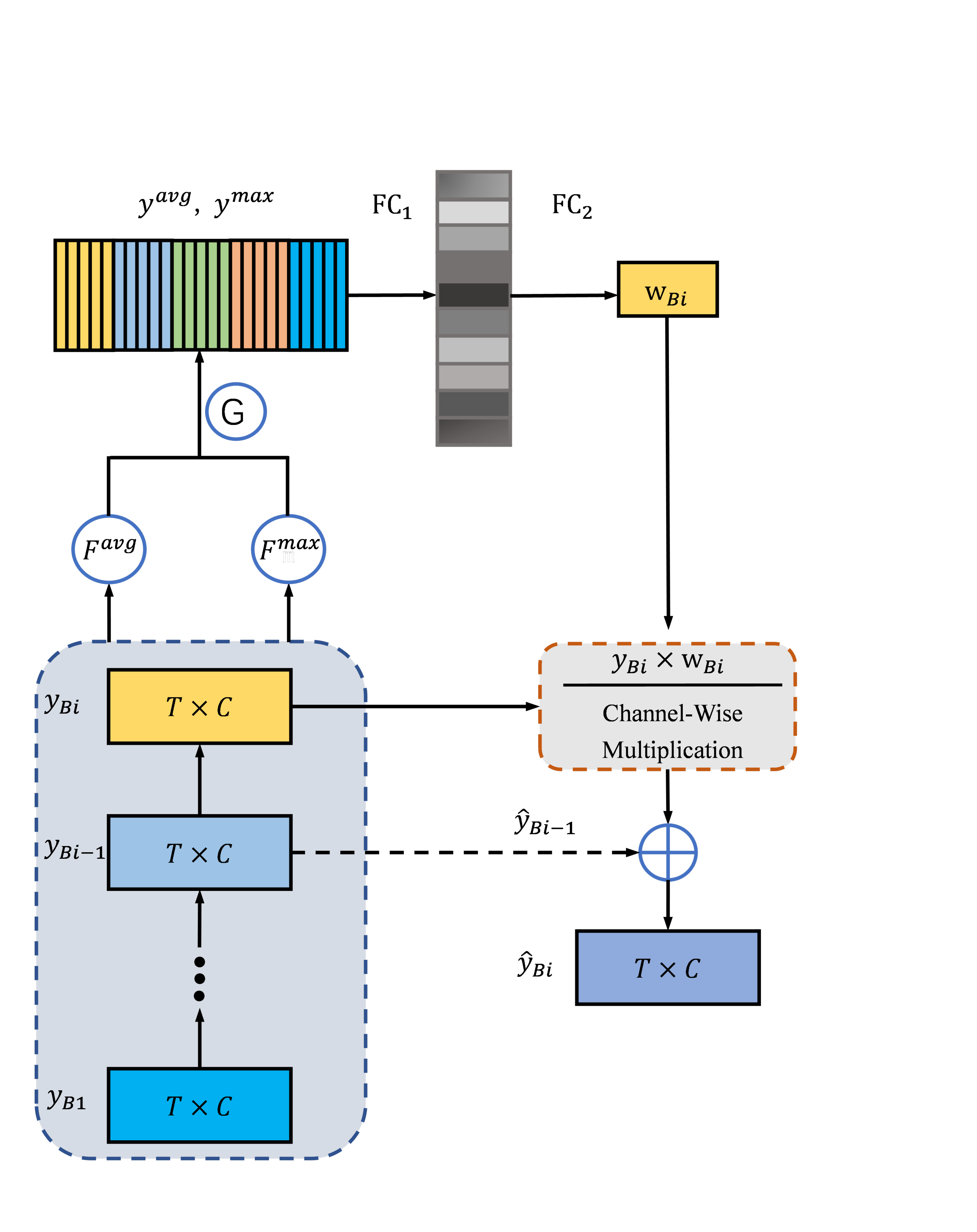}}
		\caption{Multi-Layer Feature Fusion}
		\label{fig3}
	\end{center}
\end{figure}

In deep convolutional networks, residual connections are necessary for training converge. DenseNet \cite{huang2016densely} and DenseRNet \cite{TangAcoustic} are the components of Jasper. In DenseNet, each block output was residually connected to the input of all following blocks. However, DenseNet just adds all of the residual connections together, regardless of the importance degree of each connection. In this paper, we introduce a multi-layer feature fusion module, to reweight each convolutional block by global multi-layer feature maps.

The multi-layer feature fusion module is shown as Figure~\ref{fig3}, which is inspired by the Cross-layer Feature Aggregation Module(CFA) of Cross-layer FPN \cite{li2020crosslayer}. Formally, each convolutional block output is noted as $y_{Bi}\in\mathbb{T}\times \mathbb{C}$. Here we assume the channel dimension of each block is the same for convenience. The feature fusion module firstly applies a global pooling at each block, and then concatenates all the pooling vectors to a global context $y^{avg}$ and $y^{max}$:
\begin{equation}
y^{avg} = G_{i=1}^{\mathbb{I}}(F^{avg}(y_{Bi}))
\end{equation}
\begin{equation}
y^{max} = G_{i=1}^{\mathbb{I}}(F^{max}(y_{Bi}))
\end{equation}
where $G$ is the concatenated operation across dimensions of all blocks, and $\mathbb{I}$ is the blocks number. Then we design another excitation function to compute multi-layer feature fusion weights by leveraging the context $y^{avg}$ and $y^{max}$:
\begin{equation}
\begin{split}
w_{Bi} = F^{ex}(y^{avg}, y^{max}) = \sigma(g(y^{avg}, y^{max}))\\
=\sigma(\hat{w_2}\delta(\hat{w_1}y^{avg}) \oplus \hat{w_2}\delta(\hat{w_1}y^{max}))
\end{split}
\end{equation}
where $\hat{w_1}\in\mathbb{\hat{C}}/\mathbb{\hat{D}}\times\mathbb{\hat{C}}$ and $\hat{w_2}\in\mathbb{\hat{C}}\times\mathbb{\hat{C}}/\mathbb{\hat{D}}$, are the multi-layer gating weights. $\mathbb{\hat{C}}=\mathbb{C}\times\mathbb{I}$, and $\mathbb{\hat{D}}$ is the multi-layer reduction ratio. Then each block ouput $y_{Bi}$ is reweighted by $w_{Bi}$, and add the local residual connection $\hat{y}_{Bi-1}$ to get the final block output $\hat{y}_{Bi}$:
\begin{equation}
\hat{y}_{Bi} = F^{scale}(y_{Bi}, w_{Bi}) \oplus \hat{y}_{Bi-1}
\end{equation}

\section{Experiments}
\label{sec:majhead}

\subsection{Data Set}
\label{ssec:subhead}
Our experimental works are implemented by comparing the performance of our models with original QuartzNet on Mandarin speech recognition tasks. We use AISHELL-1 \cite{bu2017aishell1} data set, a publicly available Mandarin speech corpus. AISHELL-1 contains 178 hours recording audios (16 kHz, 16 bit). The corpus is divided into training, development and testing sets. Character Error Rates (CER) are evaluated on testing set.

\begin{figure*}[t]
	\centering
	\subfigure[Multi-Resolution Attention Weight] { \label{fig4:a}
		\includegraphics[width=1.0\columnwidth]{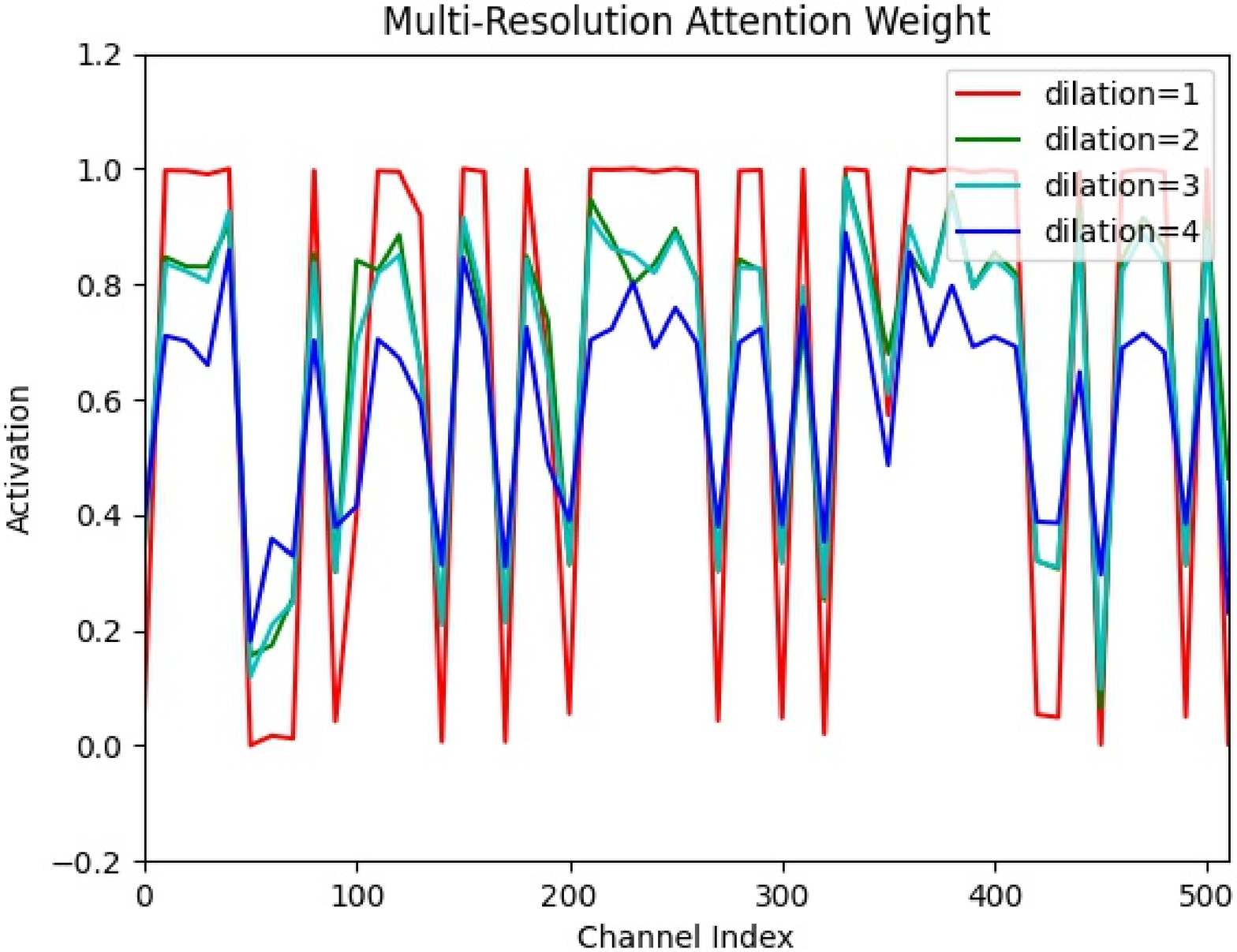}
	}
	\subfigure[Multi-Layer Feature Fusion Weight] { \label{fig4:b}
		\includegraphics[width=1.0\columnwidth]{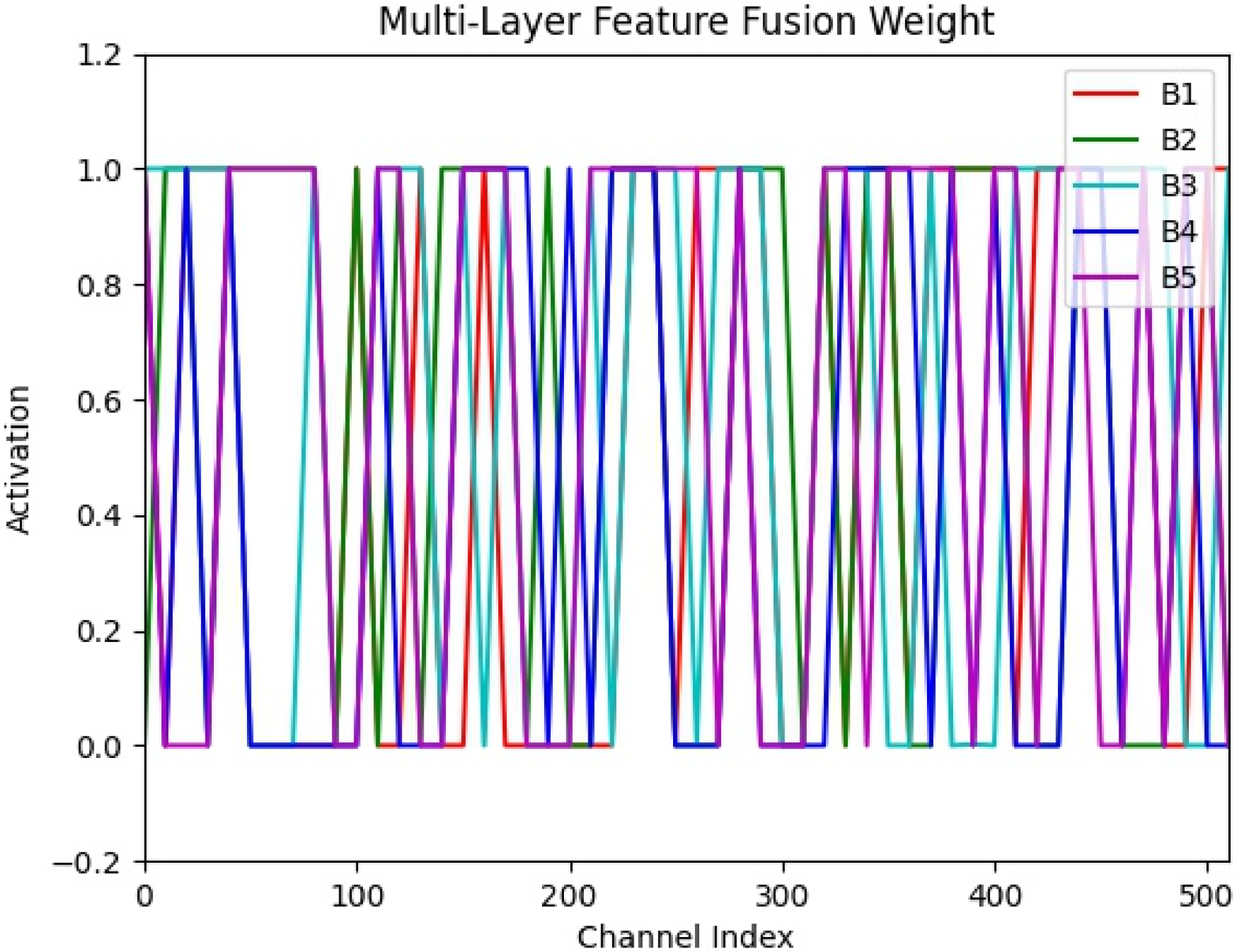}
	}
	\caption{Channel-Wise Weight Analysis}
	\label{fig4}
\end{figure*}

\subsection{Experimental Setup}
\label{ssec:subhead}

We train both small Multi-QuartzNet5x3 and large Multi-QuartzNet15x5 models on AISHELL-1 training set. The parameters of our models are listed on table~\ref{tab1} and table~\ref{tab2}. The input features are $64$-dimension mfcc coefficients with 20ms frame length and 10ms frame shift. Data augmentation ($\pm10\%$ speed perturbation, SpecAugment) are used in the training procedure. The stream numbers of multi-resolution convolution are set to $\mathbb{S}=4$ for Multi-QuartzNet5x3, and $\mathbb{S}=2$ for Multi-QuartzNet15x5 because of GPU memory limit. Reduction ratio $\mathbb{D}$ and $\mathbb{\hat{D}}$ maintain $16$ for all of the experiments. All of the models are trained for 400 epochs with batch size $32$ for each GPU. Novograd \cite{ginsburg2019stochastic} optimizer is used with learning rate of $0.01$ and weight decay of $0.0001$. We also apply the CosineAnnealing \cite{loshchilov2016sgdr} training policy with $8000$ warmup steps.

\begin{table}[h]
	\centering
	\caption{Small Multi-QuartzNet5x3 Model Configuration}
	\label{tab1}
	\begin{tabular}{p{1cm} p{0.5cm} p{0.5cm} p{0.5cm} p{1cm} p{2cm}}
		\hline
		\textbf{Block} & \textbf{$\mathbb{R}$} & \textbf{$\mathbb{M}$} & \textbf{$\mathbb{K}$} & \textbf{$\mathbb{C}$} & \textbf{Stride Set}\\
		\hline
		$C_1$ & 1 & 1 & 33 & 256 & [1]\\
		\hline
		$B_1$ & 1 & 3 & 63 & 512 & [1, 2, 3, 4]\\
		$B_2$ & 1 & 3 & 63 & 512 & [1, 2, 3, 4]\\
		$B_3$ & 1 & 3 & 75 & 512 & [1, 2, 3, 4]\\
		$B_4$ & 1 & 3 & 75 & 512 & [1, 2, 3, 4]\\
		$B_5$ & 1 & 3 & 75 & 512 & [1, 2, 3, 4]\\
		\hline
		$C_2$ & 1 & 1 & 87 & 512 & [1]\\
		$C_3$ & 1 & 1 & 1 & 1024 & [1]\\
		$C_4$ & 1 & 1 & 1 & $\lVert labels \rVert$ & [1]\\
		\hline
	\end{tabular}
	\label{table_MAP}
\end{table}

\begin{table}[h]
	\centering
	\caption{Large Multi-QuartzNet15x5 Model Configuration}
	\label{tab2}
	\begin{tabular}{p{1cm} p{0.5cm} p{0.5cm} p{0.5cm} p{1cm} p{2cm}}
		\hline
		\textbf{Block} & \textbf{$\mathbb{R}$} & \textbf{$\mathbb{M}$} & \textbf{$\mathbb{K}$} & \textbf{$\mathbb{C}$} & \textbf{Stride Set}\\
		\hline
		$C_1$ & 1 & 1 & 33 & 256 & [1]\\
		\hline
		$B_1$ & 3 & 5 & 33 & 256 & [1, 3]\\
		$B_2$ & 3 & 5 & 39 & 256 & [1, 3]\\
		$B_3$ & 3 & 5 & 51 & 512 & [1, 3]\\
		$B_4$ & 3 & 5 & 63 & 512 & [1, 3]\\
		$B_5$ & 3 & 5 & 75 & 512 & [1, 3]\\
		\hline
		$C_2$ & 1 & 1 & 87 & 512 & [1]\\
		$C_3$ & 1 & 1 & 1 & 1024 & [1]\\
		$C_4$ & 1 & 1 & 1 & $\lVert labels \rVert$ & [1]\\
		\hline
	\end{tabular}
	\label{table_MAP}
\end{table}

During inference, a standard beam search algorithm similar to \cite{Hannun2014Deep} is used with character-level 4-gram langauge model. The langauge model is trained on the transcripts of training set. Formally, $p(l|x)$ is noted as the output of Multi-QuartzNet model, $p_{lm}(l)$ is the output of language model, and $wc(l)$ is the word counts of prediction character sequence $l$. We attempt to find a sequence $l$ that maximizes the combined probability:
\begin{equation}
Q(l)=log(p(l|x))+\alpha log(p_{lm}(l))+\beta wc(l)
\end{equation}
We maximize this objective $Q(l)$ by beam search algorithm, and beam size $=200$. We also fine-tune the hyperparameter $\alpha=1.8$ and $\beta=3.5$ on the development set.

\subsection{Results}
\label{ssec:subhead}

We study the effects of each component on our proposed Multi-QuartzNet network. As illustrated in table~\ref{tab3}, for both 5x3 and 15x5 models, the performances are greatly improved with multi-resolution convolution than original QuartzNet model. With channel-wise attention module, the CER results can get a further reduction. Finally, multi-layer feature fusion enables the model to interact cross multi-layer, resulting in the best accuracy in our experiments. Our proposed Multi-Quartznet network is composed of three above components. In addition, we can see that large Multi-Quartznet15x5 model is better than 5x3 model although the Aishell-1 data set is relatively small.

\begin{table}[h]
	\centering
	\caption{The Effect of Each Component on Multi-QuartzNet, CER(\%)}
	\label{tab3}
	\begin{tabular}{p{4.1cm} p{0.8cm} p{1.1cm} p{0.8cm}}
		\hline
		\textbf{Model} & \textbf{Size} & \textbf{LM} & \textbf{Test}\\
		\hline
		\multirow{2}*{QuartzNet} & $5\times3$ & 4-gram & 8.55\\
		\cline{2-4} & $15\times5$ & 4-gram & \textbf{7.18}\\
		\hline
		\multirow{2}*{Multi-Resolution Convolution} & $5\times3$ & 4-gram & 7.79\\
		\cline{2-4} & $15\times5$ & 4-gram & \textbf{6.90}\\
		\hline
		\multirow{2}*{+Channel-Wise Attention} & $5\times3$ & 4-gram & 7.62\\
		\cline{2-4} & $15\times5$ & 4-gram & \textbf{6.84}\\
		\hline
		\multirow{2}*{+Multi-Layer Feature Fusion} & $5\times3$ & 4-gram & 7.28\\
		\cline{2-4} & $15\times5$ & 4-gram & \textbf{6.77}\\
		\hline
	\end{tabular}
	\label{table_MAP}
\end{table}
\begin{table}[h]
	\centering
	\caption{Comparsion of Multi-QuartzNet with Other Models}
	\label{tab4}
	\begin{tabular}{p{5cm} p{2cm}}
		\hline
		\textbf{Model} & \textbf{CER(\%)}\\
		\hline
		LAS \cite{Shan2019} & 10.56\\
		RNN-T \cite{Tian2019} & 11.82\\
		SA-T \cite{Tian2019} & 9.30\\
		Sync-Transformer \cite{Tian2020Synchronous} & 8.91\\
		Combiner \cite{JWu2020} & 8.87\\
		LFMMI \cite{AISHELL2017} & 7.62\\
		LDS-REG \cite{Sun2019} & 10.56\\
		ESPnet Transformer \cite{Karita2019} & 6.70 \\
		MTH-MoChA \cite{Liu2020} & 7.68 \\
		\hline
		\textbf{Multi-QuartzNet(ours)} & \textbf{6.77}\\
		\hline
	\end{tabular}
	\label{table_MAP}
\end{table}

We also compare our model with other published models in table~\ref{tab4}. Our best Multi-Quartznet15x5 model achieves CER $6.77\%$ on AISHELL-1 testing set, which outperforms most of the listed models and is close to ESPnet Transformer \cite{Karita2019}.

\subsection{Analysis}
\label{ssec:subhead}

To verify the attention and feature fusion module, we show channel-wise activations of last convolutional block $B_5$ on various dilations in Fig~\ref{fig4}a and fusion weights on different blocks in Fig~\ref{fig4}b. We observe that for small dilation $\mathbb{S}=1$, the activations tend to be closed $0$ or saturated $1$. While for big dilation $\mathbb{S}=4$, the activations are more stable, and closer to the middle region $0.5$. It demonstrates that large dilation has large reception of field, and captures more stable long-context information. While for multi-layer fusion weights, they show switch characteristics, turning on or off some channels of the outputs.

\section{Conclusion}
\label{sec:majhead}

In this paper, we propose a fully-convolutional network named Multi-QuartzNet for speech recognition. The network introduces multi-resolution convolution into the original QuartzNet. Morever, we design a channel-wise attention module and a multi-layer feature fusion module. The attention module calculates the attentions of each stream, and the fusion module computes the weights of each block. The experiments show that Multi-QuartzNet outperforms the original QuartzNet, and is close to state-of-the-art performance on AISHELL-1 data set. Future works include exploring model performances on other languages like english and experiments on larger corpus.

\section{Acknowledgement}
This paper is supported by National Key Research and Development Program of China under Grant No.2017YFB1401202, No.2018YFB0204400 and No.2018YFB1003500.

\clearpage

\bibliographystyle{IEEEbib}
\bibliography{SLT2021_Multi_Quartznet}

\end{document}